\documentclass[usenatbib,usegraphicx,letterpaper]{mn2e}
\usepackage[totalwidth=480pt,totalheight=680pt]{geometry}

\usepackage{amssymb}
\usepackage{epsfig}
\usepackage{amsmath}
\usepackage{color}

\newcommand{\unit}[1]{\mathrm{#1}}

\newcommand{\mpc}{\unit{Mpc}}

\newcommand{\kms}{\unit{km \ s^{-1}}}

\newcommand{\hmpc}{h^{-1}\mathrm{Mpc}}

\newcommand{\hkpc}{h^{-1}\mathrm{kpc}}
\newcommand{\hmsun}{h^{-1}M_{\odot}}

\newcommand{\mstar}{M_{\ast}}
\newcommand{\rvir}{R_\mathrm{vir}}

\newcommand{\msun}{\mathrm{M}_{\odot}}

\newcommand{\Omegam}{\Omega_{m}}
\newcommand{\Omegab}{\Omega_{b}}

\newcommand{\tilt}{n_{\mathrm{s}}}
\newcommand{\rhocrit}{\rho_{\mathrm{crit}}}

\newcommand{\mvir}{M_\mathrm{vir}} 
 
\newcommand{\deltavir}{\Delta_{\mathrm{vir}}}

\newcommand{\ngal}{N_{\mathrm{gal}}}
\newcommand{\nsat}{N_\mathrm{sat}}
\newcommand{\ncen}{N_\mathrm{cen}}

\newcommand{\meanm}[1]{\langle #1|\mvir \rangle}

\newcommand{\zstarve}{z_{\mathrm{starve}}}

\newcommand{\fraccqq}{\mathrm{F}_{\cqq}(\mvir)}

\newcommand{\fracsqq}{\mathrm{F}_{\sqq}(\mvir)}
\newcommand{\fracssf}{\mathrm{F}_{\ssf}(\mvir)}

\newcommand{\cqq}{C_{\mathrm{Q}}}

\newcommand{\sqq}{S_{\mathrm{Q}}}
\newcommand{\ssf}{S_{\mathrm{SF}}}

\newcommand{\ncenq}{N_{\mathrm{cen}}^{\mathrm{Q}}}
\newcommand{\nsatq}{N_{\mathrm{sat}}^{\mathrm{Q}}}
\newcommand{\ncensf}{N_{\mathrm{cen}}^{\mathrm{SF}}}
\newcommand{\nsatsf}{N_{\mathrm{sat}}^{\mathrm{SF}}}

\newcommand{\vmax}{\mathrm{V}_\mathrm{max}}

\newcommand{\vpeak}{\mathrm{V}_\mathrm{peak}}

\newcommand{\mpeak}{\mathrm{M}_\mathrm{peak}}
\newcommand{\macc}{\mathrm{M}_\mathrm{acc}}

\newcommand{\tacc}{t_{\mathrm{acc}}}

\newcommand{\beq}{\begin{equation}}
\newcommand{\eeq}{\end{equation}}
\newcommand{\beqray}{\begin{eqnarray}}
\newcommand{\eeqray}{\end{eqnarray}}

\newcommand{\ben}{\begin{enumerate}}
\newcommand{\een}{\end{enumerate}}
\newcommand{\bit}{\begin{itemize}}
\newcommand{\eit}{\end{itemize}}

\usepackage{epsfig}  \usepackage{graphicx}   \usepackage{rotating}

\begin{document}

\title[Conformity and Assembly Bias]
{Beyond Halo Mass: Galactic Conformity as a Smoking Gun of Central Galaxy Assembly Bias}


\author[Hearin, Watson \& van den Bosch]
{Andrew~P.~Hearin$^{1,2,3}$, Douglas~F.~Watson$^{2,}$\thanks{NSF Astronomy \& Astrophysics Postdoctoral Fellow} \&  Frank~C.~van den Bosch$^{4}$\\
$^1$Fermilab Center for Particle Astrophysics, Fermi National Accelerator Laboratory, Batavia, IL \\
$^2$Kavli Institute for Cosmological Physics, 5640 South Ellis Avenue, The University of Chicago, Chicago, IL \\
$^3$Yale Center for Astronomy \& Astrophysics, Yale University, New Haven, CT\\
$^4$Department of Astronomy, Yale University, P.O. Box 208101, New Haven, CT}

\maketitle

\begin{abstract}
  Quenched  central  galaxies  tend  to  reside  in  a  preferentially
  quenched large-scale environment, a  phenomenon that has been dubbed
  {\em galactic conformity}. Remarkably, this tendency persists out to
  scales far larger than the  virial radius of the halo hosting the central. Therefore,
  conformity  manifestly violates the  widely adopted  assumption that
  the  dark  matter  halo  mass  $\mvir$  exclusively  governs  galaxy
  occupation statistics. This paper is  the first in a series studying
  the  implications   of  the  observed  conformity   signal  for  the
  galaxy-dark matter  connection. We show that  recent measurements of
  conformity  on scales  $r\sim1-5$ $\mpc$  imply that  central galaxy
  quenching  statistics   cannot  be  correctly   predicted  with  the
  knowledge  of $\mvir$ alone.  We also  demonstrate that  ejected (or
  `backsplash')  satellites cannot give  rise to  the signal.  
  We then
  invoke  the   age  matching  model,  which  is   predicated  on  the
  co-evolution  of  galaxies  and  halos.   We find  that  this  model
  produces  a strong  signal,  and that  central  galaxies are  solely
  responsible.  We conclude that large-scale `2-halo' conformity represents a
  smoking  gun of {\em  central galaxy  assembly bias,}  and indicates
  that contemporary models  of satellite quenching have systematically
  over-estimated  the  influence  of post-infall  processes.
\end{abstract}




\section{INTRODUCTION}
\label{sec:intro}

The well-established connection between galaxies and dark matter halos
forms the basis of a very broad class of models of galaxy evolution
that we will loosely refer to as halo occupation models. Such models
exploit the ability of contemporary N-body simulations to calculate
the abundance, spatial distribution, and internal structure of dark
matter halos with exquisite precision. Armed with this knowledge, halo
occupation models make predictions for the observed galaxy
distribution by specifying, in a statistical sense, how galaxies
populate dark matter halos.

The Halo Occupation Distribution \citep[HOD, e.g.,][]{seljak00,
  berlind02, zheng05} and the closely related Conditional Luminosity
Function \citep[CLF, e.g.,][]{yang03, vdBosch13} are the two most
prevalent classes of halo occupation models in the literature.  In the
HOD, the galaxy-halo connection is formalized by the quantity
$P(\ngal|\mvir),$ the probability that a halo of mass $\mvir$ hosts
$\ngal$ galaxies brighter than some luminosity (or stellar mass)
threshold.  In the CLF, the quantity $\Phi(\mathrm{L}|\mvir)$ plays
the central role by specifying the mean abundance of galaxies of
luminosity $L$ found in dark matter halos of mass $\mvir.$ These
well-studied formalisms have both proven to be very powerful
theoretical tools to constrain both the galaxy-halo connection
\citep[see, for example,][and references therein]{magliocchetti03,
  yang03, zehavi05a, cooray06,zheng07, vdBosch07, zheng09, skibba_sheth09,
  simon_etal09, ross10, zehavi11, watson_powerlaw11, tinker_etal13}
and the fundamental parameters in cosmology \citep{vdBosch03cosmo,
  tinker05, leauthaud11a, more_etal13, cacciato_etal13}.

All of the above results concerning both cosmology and galaxy
evolution are predicated upon the assumption that the mass $\mvir$ of
a dark matter halo entirely determines the statistical properties of
its resident galaxy population. Yet, it is well established that the
 spatial distribution of dark matter halos depends
on halo properties besides mass \citep{gao_etal05, wechsler06,
  gao_white07, wetzel_etal07, dalal_etal08, li_etal08, lacerna11,
  lacerna12}. Here we will collectively refer to the dependence of
halo clustering upon additional halo properties besides $\mvir$ as
{\em halo assembly bias}.  Of course, if galaxy occupation statistics
depend only on $\mvir,$ then halo assembly bias only contributes
random noise in halo model predictions for the galaxy distribution.

Because halo occupation models such as the HOD and CLF have generally
been very successful at fitting observations of galaxy clustering
statistics, the possibility that the galaxy-halo connection requires
dependence on additional parameters besides $\mvir$ is generally not
considered. However, it has recently been shown in
\citet{zentner_etal13} that this assumption has the potential to
introduce significant systematic errors in halo occupation modeling of
the two-point clustering of luminosity threshold galaxy
samples. \citet{zentner_etal13} showed that these systematics can be
even more severe when color cuts comprise part of the selection
function, such as, for example, in halo occupation modeling of
color-dependent clustering \citep[e.g.,][]{zehavi11}.  In what
follows, we will generically refer to any dependence of the mapping
between galaxies and halos upon halo properties besides $\mvir$ as
{\em galaxy assembly bias}. In particular, our focus in this paper 
will be the potential correlation of galaxy color/star formation rate
with halo properties besides $\mvir.$

In the context of assembly bias, galaxy group catalogs constructed
from large redshift surveys, such as the Sloan Digital Sky Survey
\citep[SDSS:][]{york00a,DR7_09}, have proven to be particularly rich
datasets. For example, in \citet[][hereafter W06]{weinmann06b}, the authors used the
halo-based group-finding algorithm of \citet{yang_etal05} to divide
their SDSS galaxy sample into {\em central} galaxies residing at the
center of the dark matter halo of the group, and {\em satellite}
  galaxies orbiting around the central within the potential well of
the group's halo. W06 found that both
the $g-r$ color and star formation rate (SFR) of satellite galaxies
depends on the color/SFR of the group's central galaxy {at fixed halo
  mass},\footnote{Of course the true dark matter halo mass of a galaxy
  group is not directly observable, and so in W06 the authors use
  total group luminosity as their halo mass proxy. The role played by
  this choice and other details associated with this conformity
  measurement will soon appear in Campbell et al., in prep.} a
phenomenon the authors dubbed {\em galactic conformity}.\footnote{See
  also \citet{phillips_etal14} for measurements of this phenomenon in
  a sample of satellites of isolated Milky Way-mass galaxies.} If
correct, this observation manifestly violates the assumption that
galaxy assembly bias is zero, as the properties of the satellites were
explicitly shown to have an additional dependence upon some property
besides $\mvir$ (in particular, satellite color/SFR evidently also
depends on the color/SFR of the central).

In a recent, closely related paper, \citet[][hereafter K13]{kauffmann_etal13} 
used SDSS data to 
demonstrate that correlations between star
formation indicators of central galaxies and their neighboring
galaxies persist out to several $\mpc,$ far outside the virial radius
of the host halos of the centrals. Although K13 also used the term
``conformity'' to refer to the signal they measured, from the
perspective of the halo model \citep[e.g.,][]{Cooray02,mo_vdb_white10} the W06 and
K13 measurements are qualitatively distinct: the former refer to SFR
correlations between galaxies occupying the same dark matter halo,
while the latter considers galaxies in distinct halos. Hence, in what
follows we refer to the conformity signals detected by W06 and K13 as 
``1-halo'' and ``2-halo'' conformity, respectively.

This paper is the first in a series investigating the implications of
galactic conformity for halo occupation statistics and for the physics
of galaxy formation (in particular galaxy quenching). In this paper we argue
that {\em central} galaxy assembly bias is required for any
galaxy-halo model to produce non-zero 2-halo conformity. In a
companion paper to the present work (Paper II), we will show that
1-halo conformity can naturally be encoded in a generalized HOD
formalism, and demonstrate that the W06 measurements indicate that
this signal is strong enough to significantly impact small-scale
clustering. Each of these two ``theory papers'' will be accompanied by
its own follow-up paper providing new measurements of the
corresponding conformity signal. In the observational follow-up to the
present paper (Paper III), we will quantitatively compare our updated
K13 measurements to predictions from both empirical and semi-analytic
models (SAMs) of galaxy formation, and seek to identify the
ingredients necessary to bring theoretical predictions into accord
with the observations. Finally, in Paper IV, we will conduct a
likelihood analysis using the generalized HOD model developed in Paper
II to provide quantitative constraints on the co-evolution of central
and satellite galaxies, as well as other signatures of galaxy
evolution that have been previously neglected in HOD modeling of
SFR-dependent galaxy clustering.

This paper is organized as follows. We provide an overview of the
conformity measurement in \S\ref{sec:twohaloconf}.  In
\S\ref{sec:halomodel} we outline the various formulations of the
galaxy-halo connection we use to model the galaxy distribution. Our
primary results are presented in \S\ref{sec:twohaloconfmockobs}, and
we discuss the implications of our findings in
\S\ref{sec:discussion}.  Throughout the paper we assume a flat
$\Lambda$CDM cosmological model with $\Omega_{\mathrm{m}}=0.27$ and
Hubble constant $H_0=70$ km s$^{-1}$ Mpc$^{-1}$.


\section{2-HALO CONFORMITY}
\label{sec:twohaloconf}

\subsection{Definition of the Signal}
\label{subsubsec:twohalodef}

As outlined in \S\ref{sec:intro}, K13 recently demonstrated that star
formation indicators of central galaxies and their neighbors are
correlated out to several $\mpc$. 
From a volume-limited galaxy sample constructed from 
the New York University Value-Added Catalog \citep{VAGC_05}, 
K13 first select galaxies in a specific range
of stellar mass $\mstar$. For a galaxy to be identified as a
central, it was required that no other galaxy with stellar mass greater
than $\mstar/2$ lie within a projected distance of $500$ kpc and a
velocity difference of $500$ km/s of the candidate central. Throughout
this paper, we will refer to any galaxy that passes these isolation
criteria as an ``isolated primary''. We will reserve the term
``central'' for galaxies that truly reside at the centers of host
halos.

After applying the above isolation criteria, it was shown in K13 that
when a sample of isolated primaries of the same stellar mass is
divided into subsamples according to SFR, the mean specific star
formation rate (defined as $\langle\mathrm{SFR}/\mstar\rangle$, and
written as $\langle\mathrm{sSFR}\rangle$) of galaxies neighboring
quenched primaries is suppressed relative to the neighbors of
primaries that are actively forming stars. As can be seen in Figs. 2
$\&$ 3 of K13, the magnitude of the effect is quite strong, and
measured with high statistical significance. Moreover, for a sample of
primaries with $\mstar\sim10^{10}\msun,$ K13 found that the difference
in $\langle \mathrm{sSFR} \rangle$ between the samples of neighbors
persists out to at least $4\mpc,$ {\em well beyond the virial radius
  of the host halo of the primary}. As already mentioned above, to
distinguish this conformity signal from that introduced and identified
by W06, we refer to this large-scale correlation between sSFRs as
``2-halo conformity''.

\subsection{Significance for Halo Occupation Models of Galaxy Quenching}
\label{subsubsec:hodsig}

Non-zero 2-halo conformity represents a clear violation of the core
assumption in virtually all halo occupation models, namely that galaxy
properties are, in a statistical sense, governed by halo mass alone.
Although a violation of this `$\mvir$-only' ansatz is expected for
some galaxy properties, such as galaxy size (which is believed to
be strongly related to the halo's angular momentum: \citealt{mo_mao_white98,kravtsov13}), numerous studies
have assumed that galaxy SFR (and hence, color) can still be modeled
under the $\mvir$-only assumption \citep[e.g.,][]{zehavi05a,zehavi11,collister_lahav05,skibba_sheth09,tinker_etal13,guo_etal14}. 
Non-zero conformity
not only violates one of the core assumptions on which these and other
studies are based, but also has important implications for the physics
of galaxy quenching. In particular, taken at face value, 2-halo
conformity suggests that large-scale environment in addition to halo mass has an important
impact on controlling the SFRs of individual galaxies.

However, before concluding that observed quenching statistics violate the $\mvir$-only ansatz, 
it is important to verify
that the conformity signal detected by K13 is not a mere manifestation
of observational systematics and/or sample selection effects.  For
example, no isolation criterion is perfect, and so any sample of
isolated primaries will be contaminated at some level by galaxies that
are truly satellites. Such contamination has the potential to
masquerade as 2-halo conformity for the following reason. At fixed
stellar mass, it is by now well-established that satellites have a
larger quenched fraction than centrals \citep[e.g.,][]{vdBosch08,
  Wetzel_Tinker_Conroy12, watson_etal14}.  This implies that (1) when a
sample of isolated primaries is divided into quenched and star-forming
sub-populations, the quenched subsample will have a relatively larger
fraction of contaminating satellites.  Now, satellites tend to reside
in more massive dark matter halos than centrals of the same stellar
mass \citep{yang08}; halo clustering strength increases with halo mass
\citep{MoWhite96}; and the quenched fraction of galaxies increases
with halo mass (W06). Therefore: (2) the large-scale environment
surrounding a satellite galaxy will tend to be more quenched than the
environment of a central of the same stellar mass. Putting (1) and (2)
together, satellite contamination of the isolated primary sample
creates the potential to erroneously conclude that quenched isolated
primaries reside in a preferentially quenched large-scale environment,
even when no signal around true centrals is present.

In this paper we use different mock galaxy catalogs, described in
detail in \S\ref{sec:halomodel} below, to investigate whether this,
and other potential systematics of the K13 measurements, 
can result in a false signal of 2-halo conformity. The
mocks we use represent realizations of different models of galaxy
quenching, permitting us to directly investigate the characteristic
features of halo occupation statistics that can, and cannot, give rise
to the 2-halo conformity signal as measured in K13.


\section{HALO OCCUPATION MODELS}
\label{sec:halomodel}

In this section we describe the collection of halo occupation models,
and their corresponding mock catalogs, that we use in our
investigation of the 2-halo conformity signal. These mocks are all
built by populating dark matter halos in the high-resolution Bolshoi
simulation, described in \S\ref{subsec:sim}. In each case, the mock
galaxies are assigned stellar masses using standard
(sub)halo abundance matching (SHAM) as outlined in
\S\ref{subsec:stellarmass}.  The models differ markedly, though, in
how they connect SFRs to the mock galaxies, and so we
separately review each model's approach in
\S\ref{subsec:hodquenching}.  For convenience, we provide a brief
summary of our mocks in \S\ref{subsec:mocksummary}. We conclude this
section in \S\ref{subsec:modelcomparison} by comparing the halo
occupation distributions and two-point correlation functions of the
mocks, explicitly demonstrating that these models give broadly similar
predictions for these traditional statistics despite the radically
different assumptions upon which the models are based.

\subsection{Simulation and Halo Catalogs}
\label{subsec:sim}

The basis of the mock galaxy catalogs we construct in this paper is
the collisionless N-body Bolshoi simulation \citep{bolshoi_11}. The
cosmological parameters of the Bolshoi run are $\Omegam=0.27,$
$\Omega_{\Lambda}=0.73,$ $\Omegab=0.042,$ $\tilt=0.95,$ $\sigma_8 =
0.82,$ and $H_0=70$ $\kms\mpc^{-1}.$ The simulation uses the Adaptive
Refinement Tree code \citep[ART;][]{kravtsov_eta97,
  gottloeber_klypin08} to solve for the evolution of $2048^3$
particles in a $250\hmpc$ periodic box. Each particle has a mass of
$\mathrm{m_p}\approx1.9\times10^{8}\hmsun;$ the force resolution of
the simulation is $\epsilon\approx1\hkpc.$ Bolshoi snapshot data and
halo catalogs are part of the Multidark Database \citep{riebe_etal11},
available at {\tt http:www.multidark.org}.

The catalogs of dark matter halos that we use to populate our mocks
are publicly available at {\tt
  http://hipacc.ucsc.edu/Bolshoi/MergerTrees.html} and have been
obtained using the (sub)-halo finder ROCKSTAR
\citep{behroozi_rockstar11, behroozi_trees13}, which uses adaptive
hierarchical refinement of friends-of-friends groups in 6
phase-space dimensions and one time dimension. Halos in these catalogs
are defined to be spherical regions centered on a local density peak,
such that the average density inside the sphere is
$\deltavir\approx360$ times the mean matter density of the simulation
box. The radius of each such sphere defines the {virial radius}
$\rvir$ of the halo, which is related to the mass of the halo via
$\mvir=\frac{4}{3}\pi\rvir^3\deltavir\Omegam\rhocrit,$ where
$\rhocrit=3H_{0}^{2}/8\pi\mathrm{G}$ is the critical energy density of
the universe. Subhalos in this catalog are distinct, self-bound
structures that are found within the virial radius of a larger
``host'' halo.

\subsection{Modeling the $\mathrm{M}_{*}$-Halo Connection} 
\label{subsec:stellarmass}

The first step of constructing a mock galaxy distribution is assigning
galaxies with stellar mass, $M_*$, to dark matter halos or
subhalos in the Bolshoi simulation. In all cases considered here, this
is done using the popular SHAM technique
\citep{kravtsov04a, vale_ostriker04, tasitsiomi_etal04, conroy06,
  shankar_etal06, trujillo_gomez11, rod_puebla12,
  watson_etal12b}. Abundance matching proceeds by presuming that the
stellar mass of the galaxy occupying a halo increases monotonically
with some (sub)halo property, $x$. For a given amount of scatter
between $\mstar$ and $x$, there exists a unique mapping from $x$ to
$\mstar$ that reproduces the observed stellar mass function.  In all
but one of the models studied in this paper, we use the halo property
$x = \vpeak$, the largest value of the maximum circular velocity
$\vmax$ that the halo ever attains through its entire assembly
history. We refer the reader to \citet{hearin_etal13b} for a
description of our volume-limited galaxy
catalog, and Appendix~A of \citet{hearin_etal12b} for details
concerning our SHAM implementation.  Additionally, we
use a mock catalog kindly provided to us by Andrew Wetzel, in which
the abundance matching is instead performed using $x = \mpeak$, the
largest value of the virial mass ever attained by the halo \cite[for
  details, see][]{Wetzel_Tinker_Conroy12}. All mocks considered
here are composed of galaxies with $\mstar > 10^{9.8}\msun$, 
and include scatter of $0.15$ dex between stellar mass and
the chosen halo property, in rough agreement with observational
constraints \citep[e.g.,][]{cooray06, yang09a, more10}.

We emphasize that, despite its striking simplicity, SHAM has been
shown to yield a rich variety of statistics of the galaxy
distribution that are in good agreement with observations, including
two-point clustering over a wide range of
redshifts \citep{conroy06}, the CLF of galaxy groups \citep{reddick12},
magnitude gap statistics \citep{hearin_etal12b}, galaxy-galaxy lensing
\citep{hearin_etal13b}, and, indirectly, the galaxy size-stellar mass
relation \citep{kravtsov13}.

\subsection{Modeling the SFR-Halo Connection} 
\label{subsec:hodquenching}

Halo occupation models can also be employed to characterize how the
SFR of a galaxy is connected to the properties of its halo. Two of the
models we study allow for the ability to predict the full,
continuously valued SFR distribution of mock galaxies. However, in one
of our models it is only possible to divide the mock galaxy sample of
interest into ``star-forming'' and ``quenched'' sub-populations. To
make the predictions of these models commensurable, we designate a
galaxy as quenched when $\mathrm{sSFR} < 10^{-11}$
$\mathrm{yr}^{-1}.$ This cut is chosen to be in accord with the
``persistent bimodality'' of the observed SFR distribution
\citep{Wetzel_Tinker_Conroy12}.

\subsubsection{Standard HOD Model}
\label{subsubsec:tradhodquenching}

As mentioned in \S\ref{sec:intro}, in the HOD formalism, the central
quantity is $P(\ngal|\mvir).$ This quantity is typically decomposed
into central and satellite contributions, accomplished by writing the
first moment of $P(\ngal|\mvir)$ as $\meanm{\ngal} = \meanm{\ncen} +
\meanm{\nsat}$. It is common convention to then independently
parameterize the functional forms of the central and satellite moments
\citep[see, e.g.,][and references therein]{zheng07}. In our SHAM-based mocks,
central galaxies are galaxies that occupy distinct host halos, while
satellite galaxies occupy subhalos; hence, satellite galaxies are in
orbit around a central galaxy.

It is straightforward to extend this technique to further predict how
quenched and star-forming subpopulations occupy dark matter halos: one
simply allows the HOD parameters of star-forming and quenched
populations to take on independent values \citep[e.g.,][]{zehavi11}.
Alternatively, one may choose to model the stellar mass-dependence of
galaxy occupation statistics with a standard halo model, and
additionally specify the fraction of centrals and satellites in a halo
of mass $\mvir$ that are quenched, $\fraccqq$ and $\fracsqq,$
respectively \citep[see][for an early example]{vdBosch03}. We adopt
this second approach for the first mock catalog used in the present
paper, assuming log-linear forms for the quenched fractions of
satellites
\beq \label{eq:satellitequenching}
\fracsqq =  A_{\mathrm{sat}} + \left[\log_{10}(\mvir) - 11\right]B_{\mathrm{sat}}\,, 
\eeq 
and centrals
\beq \label{eq:centralquenching}
\fraccqq = A_{\mathrm{cen}} + \left[\log_{10}(\mvir) - 11\right]B_{\mathrm{cen}}\,.
\eeq 
Here $A_{\mathrm{x}}$ and $B_{\mathrm{x}}$ are parameters of the model
that in general depend on the stellar mass threshold of the galaxy
sample.  Note that both $\fracsqq$ and $\fraccqq$ must be bounded by
zero and unity at all $\mvir$, which we implement through the use of
$\tilde{\mathrm{F}}_{x},$ defined as
\beq
\tilde{\mathrm{F}}_{x}(\mvir) \equiv \mathrm{MAX}
\left[0,\mathrm{MIN}[\mathrm{F}_{x}(\mvir),1]\right]\,.
\eeq
Observations of both galaxy group statistics \citep{weinmann06b,
  hearin_etal13b, watson_etal14} and two-point clustering
\citep{vdBosch03, ross_brunner09, tinker_etal13} suggest that these
functions increase monotonically with halo mass, so that
$B_{\mathrm{sat}}>0$ and $B_{\mathrm{cen}}>0.$

In this formulation, the average numbers of quenched and star-forming
satellite galaxies in a halo of mass $\mvir$ are given by
\beqray\label{eq:nsatqdef}
\meanm{\nsatq} & = & \fracsqq\meanm{\nsat} \\ \nonumber
\meanm{\nsatsf} & = & \fracssf\meanm{\nsat}, 
\eeqray 
respectively, where $\fracsqq + \fracssf = 1$. Directly analogous
expressions apply to the corresponding occupation numbers of centrals;
$\meanm{\ncenq}$ and $\meanm{\ncensf}$.
 
For the fiducial values of the parameters governing the quenched
fractions in this model, we choose $A_{\mathrm{cen}} = 0.2,$
$B_{\mathrm{cen}} = 0.15,$ $A_{\mathrm{sat}} = 0.25,$ and
$B_{\mathrm{sat}} = 0.15.$ These fiducial values were chosen based on
the corresponding values predicted by the age matching model,
described below in \S\ref{subsubsec:agematching}. Since the age
matching model has already been shown to make highly realistic
predictions for the galaxy distribution observed in SDSS, tuning our
HOD parameters to resemble the age matching model's ensures that the
resulting mock is realistic in turn. However, we stress that all of
our conclusions regarding 2-halo conformity are insensitive to this choice.

Once a set of fiducial values are chosen for the HOD parameters, it is
straightforward to construct a mock galaxy catalog that serves as a
realization of the model. As described in \S\ref{subsec:stellarmass},
we begin with abundance matching to create a sample of mock centrals
and satellites.  Next we assign each galaxy as either `quenched' or
`star forming' using simple Monte Carlo techniques, where we interpret
$\fraccqq$ and $\fracsqq$ as the probabilities that a central or
satellite galaxy is quenched.


\begin{figure*}
\begin{center}
\includegraphics[width=7cm]{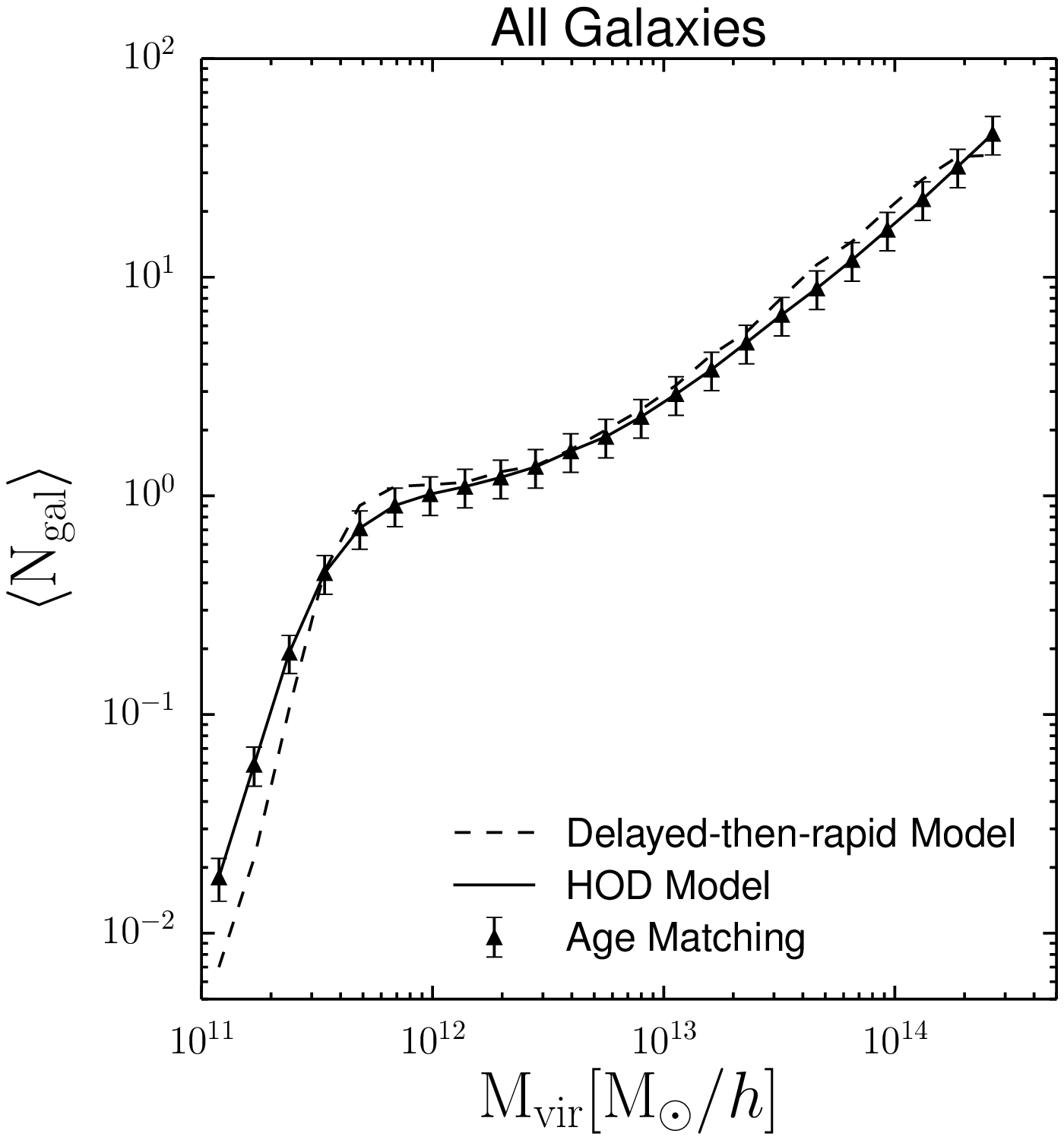}
\includegraphics[width=7cm]{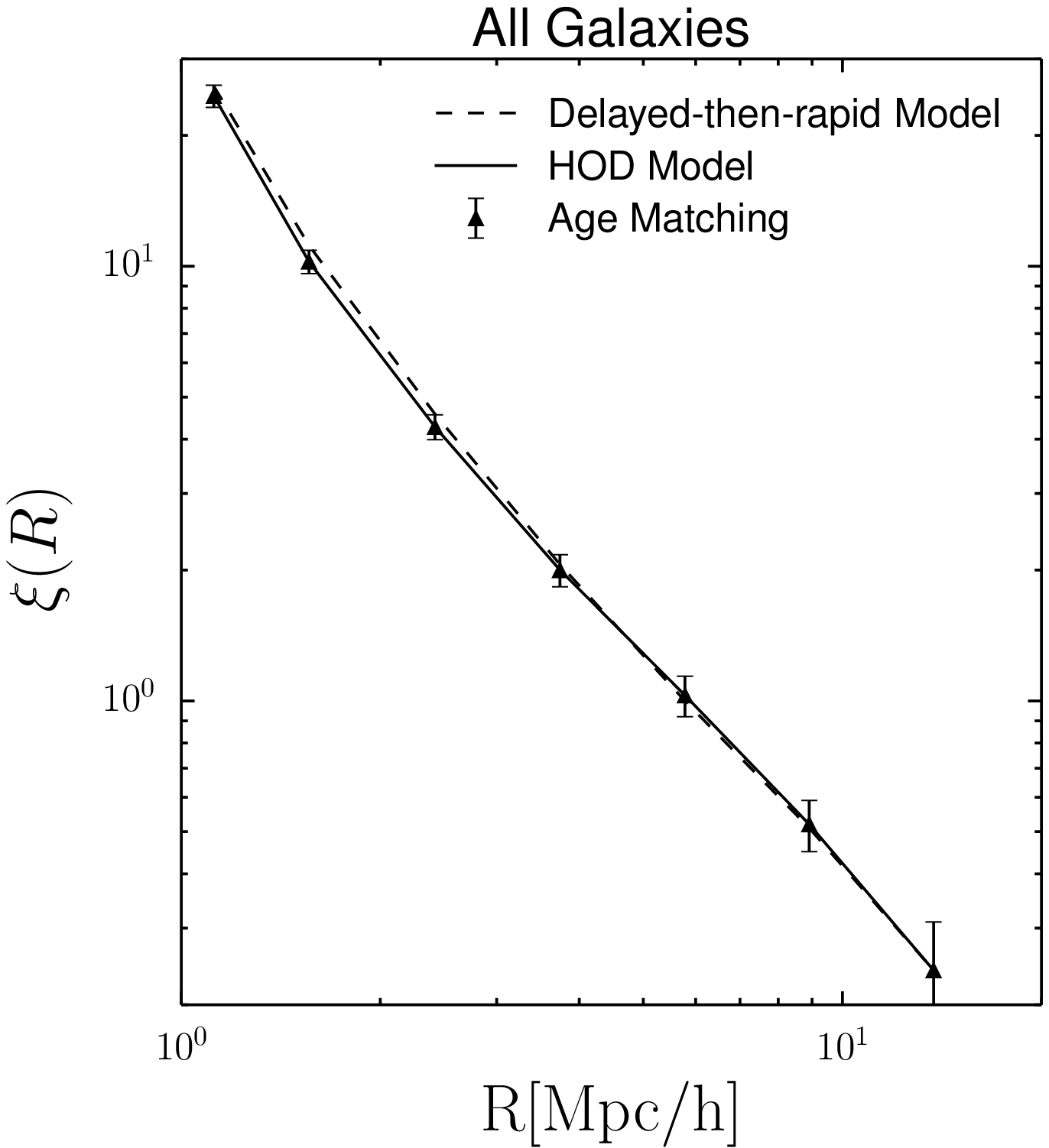}
\includegraphics[width=7cm]{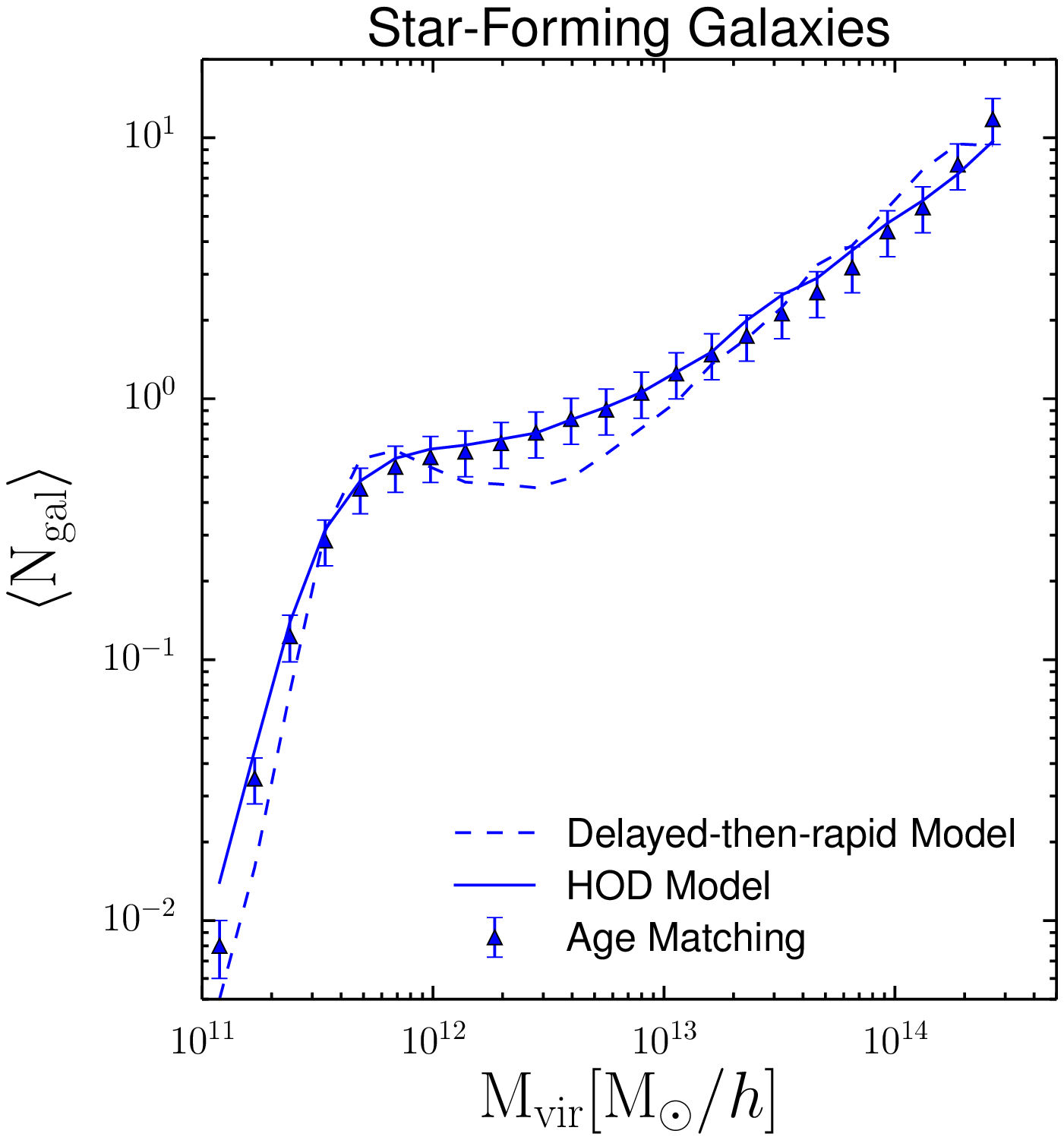}
\includegraphics[width=7cm]{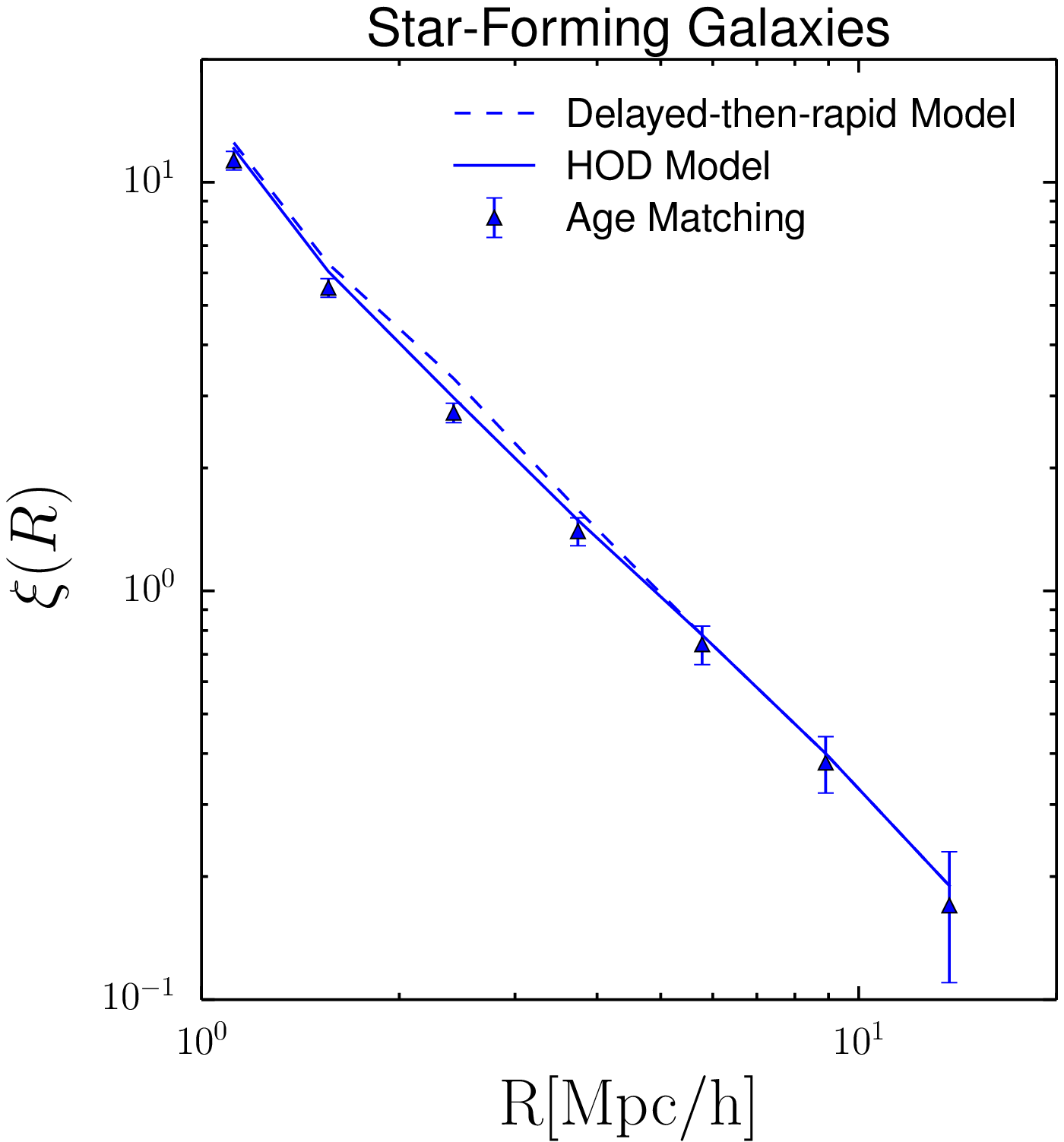}
\includegraphics[width=7cm]{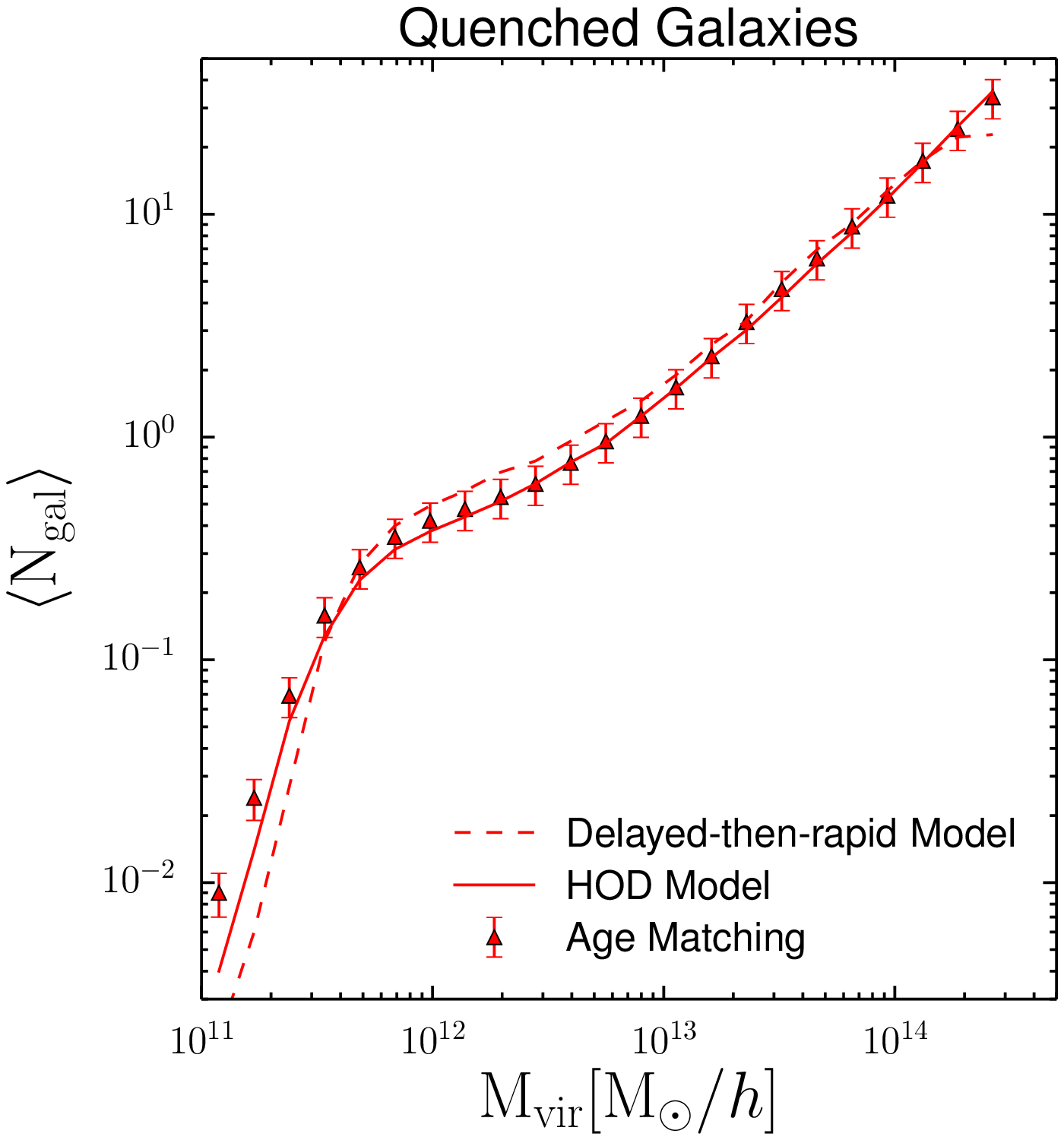}
\includegraphics[width=7cm]{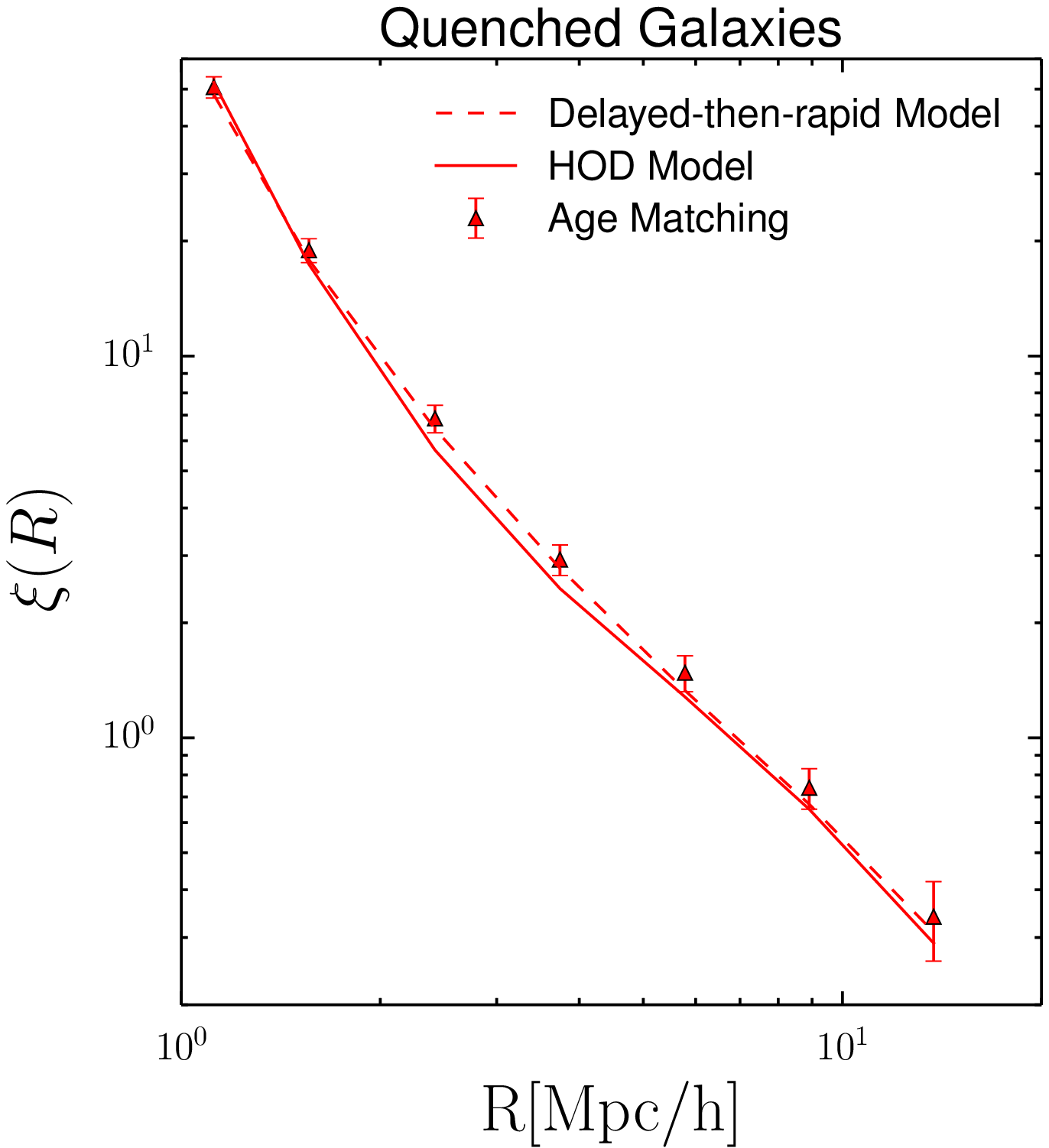}
\caption{{\bf Comparison of our three quenching models.} The {\em
    left} column of panels shows the mean number of galaxies
  occupying halos of a given mass $\mvir.$ 
  Two-point correlation functions appear in the 
  {\em right} panels, plotted as a function of 3D separation.  
  The top panels refer to the full, volume-limited mock
  galaxy samples (with stellar mass thresholds $\mstar >
  10^{9.8}\msun$) in each mock, the middle and bottom panels
  show the star-forming and quenched subpopulations, respectively.  The broad
  agreement between these galaxy distributions is striking given how
  radically these models differ from one another in their predictions
  for the processes that govern quenching. See \S\ref{subsec:newprobe} for further discussion.}
\label{fig:modelcomparisons}
\end{center}
\end{figure*}


\subsubsection{Delayed-Then-Rapid Model}
\label{subsubsec:wetzelmodel}

The SFR designation in the second mock catalog comes from the
``delayed-then-rapid'' quenching model introduced in
\citet{Wetzel_Tinker_Conroy12}, to which we
refer the reader for details. Briefly, a mock central galaxy with stellar mass $\mstar$ is
assigned an SFR value by randomly selecting an SDSS central galaxy
with a similar $\mstar,$ where the group finding algorithm introduced
in \citet{tinker_wetzel11} is used to identify SDSS centrals.
Because abundance matching was used to assign the value of $\mstar$ to
host halos in this catalog, this model effectively assumes that
central galaxy quenching statistics are governed by $\mvir$ alone.

For satellite galaxies, which are associated with present-day
subhalos, special physical significance is attached to $\tacc$, the
epoch the subhalo first passes within the virial radius of some other,
more massive host halo. Prior to $\tacc$, satellite galaxies are
presumed to have evolved as centrals. Using a model for how the SFR
distribution of central galaxies scales with redshift, each
present-day satellite galaxy is first assigned a SFR at time $\tacc$
that is appropriate for its stellar mass at that epoch. If at
accretion the satellite is already quenched, it is assumed to remain
quenched. Otherwise, its SFR is assumed to undergo exponential
quenching, on an e-folding time scale of $\tau_{\rm Q} = \tau_{\rm
  Q}(M_*) \simeq 0.2-0.8$Gyr, but only after a delay-time of $t_{\rm
  delay} = t_{\rm delay}(M_*) \simeq 2 - 4$Gyr. The stellar mass
dependencies of $\tau_{\rm Q}(M_*)$ and $t_{\rm delay}(M_*)$ have been
tuned to match the quenched fractions and quenching gradients of
satellite galaxies in the \citet{tinker_wetzel11} SDSS group catalog.
Thus the SFRs of satellites in this model are determined by both halo
mass {\it at} accretion, $\macc$, and by the time {\it since}
accretion, $t-\tacc$.

As shown in \citet{wetzel_etal13}, a significant sub-population of
present day host halos is comprised of ``ejected'' (or ``backsplash'') galaxies, which are halos that have
previously been identified as subhalos at some point in their assembly
history. These objects are modeled in an identical fashion as the
satellites associated with present-day subhalos, so that ejected
satellites quench exponentially at a time $t_{\rm delay}$ after they
first passed within a distance $\rvir$ of some larger halo. As shown in
\citet{wetzel_etal13}, treating backsplash galaxies on par with
satellite galaxies provides an explanation for the enhanced SFR quenching
observed out to $\sim 2.5 \rvir$ for groups and clusters. 
 However, as we show below, backsplashing is not sufficient to explain the
2-halo conformity observed by K13.

\subsubsection{Age Matching Model}
\label{subsubsec:agematching}

The third mock catalog used in this paper is generated with the age
matching technique.  The formalism was introduced in \citet{HW13A} and
extended in \citet{hearin_etal13b}, and was shown to predict a variety
of SDSS galaxy statistics as a function of $g - r$ color.  It was
recently generalized to predict SFRs in
\citet{watson_etal14}. In age matching, no explicit distinction is
made between centrals and satellites, and no special significance is
attached to $\rvir.$ At fixed stellar mass, SFR values are randomly
drawn from the distribution of SDSS galaxies with the corresponding
$\mstar.$ The lowest SFR values are assigned to the ``oldest''
(sub)halos, where (sub)halo age is quantified by the halo property
$\zstarve,$ which is primarily determined by the epoch in the very
distant past where the (sub)halo transitioned from the fast- to
slow-mass accretion regime \citep{wechsler02,zhao03}. Thus in age
matching, galaxy assembly bias of both centrals and satellites is
quite strong, since at fixed $\mstar$ galaxy SFR is in monotonic
correspondence with a marker of halo assembly time \citep[see][]{zentner_etal13}.

\subsection{Summary of Mocks}
\label{subsec:mocksummary}

\bit
\item[] {\bf Mock 1:} {\em Standard HOD model.} Quenching of both
  centrals and satellites depends only on $\mvir.$
\item[] {\bf Mock 2:} {\em Delayed-then-rapid model.} Quenching of
  centrals depends only on $\mvir.$ Quenching of satellites and
  backsplash galaxies depends on both $\macc$ and $(t-\tacc)$.
\item[] {\bf Mock 3:} {\em Age matching model.} Quenching of centrals
  and satellites depends on both $\mvir$ and (sub)halo formation time. 
 \eit

\subsection{Model Comparison}
\label{subsec:modelcomparison}

We conclude this section by comparing some statistics of the three
mocks introduced above.  For mock
galaxies with $\mstar > 10^{9.8} \msun,$ 
Fig.~\ref{fig:modelcomparisons} shows the
HODs, $\meanm{\ngal}$, in the left-hand panels, and two-point
correlation functions, $\xi(r)$, in the right-hand panels. Results are shown for all
galaxies (upper panels), star-forming galaxies (middle panels), and
quenched galaxies (lower panels).

All three mocks have, by construction, the same stellar mass function,
but differ in the way the mock galaxies were split into quenched and
star forming sub-populations. Although none of the mocks has been
tuned to reproduce the observed clustering, or to agree
with the clustering of any of the other mocks, they have halo
occupation statistics and clustering properties that are remarkably
similar. This indicates that both of these one- and two-point
functions, which are commonly used to quantify halo occupation
statistics, are largely insensitive to the differences between these
three mock galaxy distributions.  This is despite the fact that they
represent radically different perspectives on the physical processes
that drive galaxy quenching. These results are especially noteworthy
because, as we will see, 2-halo conformity brings the differences
between these models into sharp relief.


\begin{figure}
\begin{center}
\includegraphics[width=8cm]{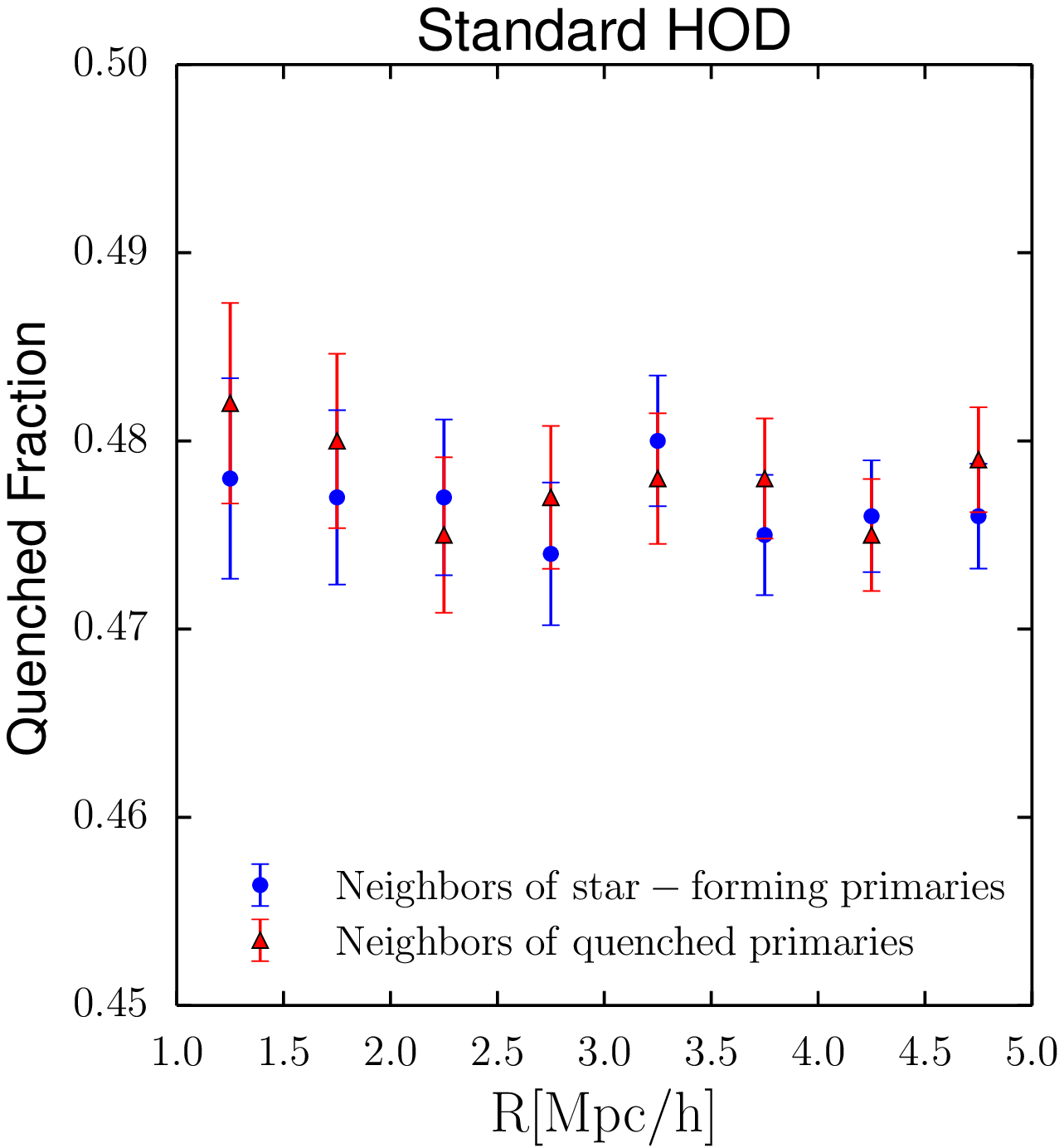}
\includegraphics[width=8cm]{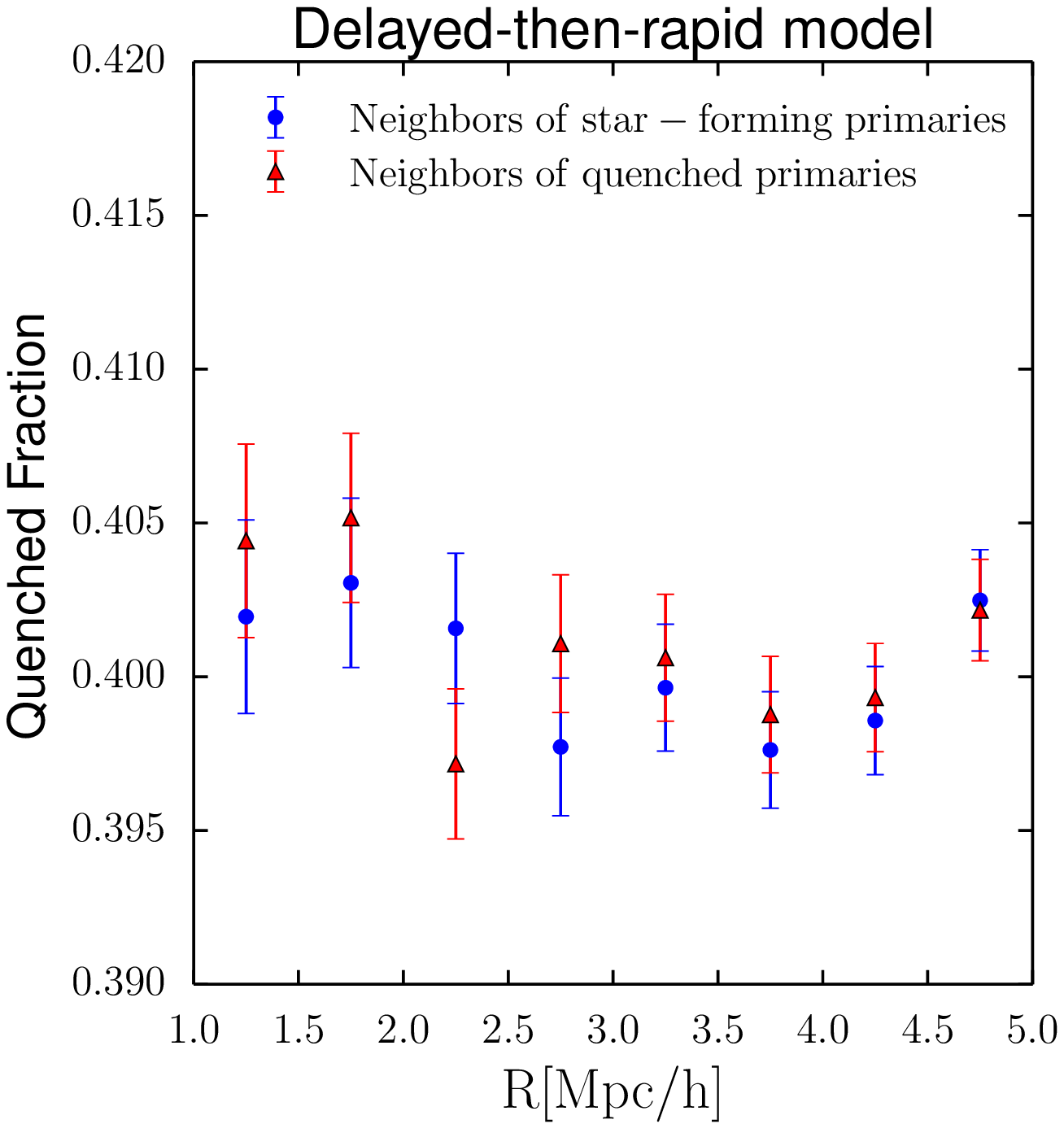}
\caption{ In each panel, the vertical axis shows the mean quenched
  fraction of all galaxies neighboring samples of ``isolated
  primaries'' selected in the same manner as in
  \citet{kauffmann_etal13}. Horizontal axes are the 3D distance from the primary.  
  The {\em top} panel shows the 2-halo conformity signal around
  isolated primaries with $10^{10}\msun < \mstar < 10^{10.5}\msun$ as
  predicted by the standard HOD quenching model described in
  \S\ref{subsubsec:tradhodquenching}. Standard HOD models predict that
  the SFR of galaxies occupying distinct halos are uncorrelated,
  giving rise to zero 2-halo conformity, in contrast to the relatively
  strong signal measured in K13 (see their Figs. 2 $\&$ 3).  The {\em
    bottom} panel is the same as the top panel, only here we show the
  prediction of the ``delayed-then-rapid'' model described in
  \S\ref{subsubsec:wetzelmodel}. There is little-to-no signal, again
  in contrast with the K13 measurements. This demonstrates that
  satellite backsplashing alone cannot account for the observed level
  of 2-halo conformity, and furthermore implies that the role of
  post-infall processes on satellite quenching has been over-estimated
  in this and related models. See \S\ref{subsec:wetzeltwohalo} and
  \S\ref{subsec:thenewjam} for further discussion. }
\label{fig:hodcentralconformity}
\end{center}
\end{figure}


\section{Mock Observations of 2-Halo Conformity}
\label{sec:twohaloconfmockobs}

We now proceed to apply the K13 methodology to our mocks in order to
investigate whether they reveal any sign of 2-halo conformity.  We begin 
by identifying isolated primary galaxies in the mocks using the K13 criteria. 
In particular, we place the mock galaxies into redshift-space by invoking
 the distant observer approximation, using the simulation's z-axis as the 
 line-of-sight. A galaxy with stellar mass $\mstar$ is labeled as an isolated primary 
 if a cylinder with a line-of-sight length of $\pm 500$ km/s and radius of $500$ kpc/h contains no other 
 mock galaxies with stellar mass greater than $\mstar/2.$ 
  
We deviate from K13 in that we measure 2-halo
conformity using the mean quenched fractions, rather than $\langle \mathrm{SFR} \rangle$. As discussed in
\S\ref{subsec:hodquenching}, this choice is made to simultaneously accommodate all
three of our mocks. In addition, whereas K13 measured 2-halo
conformity in projection, and presented the signal as a function of
the radius of the cylindrical annulus centered on the isolated
primary, we present our measurements of 2-halo conformity as a
function of the 3D distance from the isolated primary.  Since the
salient point of this paper is that halo occupation models without
central galaxy assembly bias contain no statistically significant
level of 2-halo conformity, it suffices to demonstrate that this fact
holds true when the signal is measured in the 3D galaxy
distribution. After all, projection effects will only diminish the
true signal strength. Finally, for brevity, we only focus on the
stellar mass range $10^{10}\msun < \mstar < 10^{10.5}\msun$ of
primaries, which corresponds to the lower end of the range studied in
K13, for which the observed signal is strongest. In the Appendix, we
investigate how the signal varies with the stellar mass of the
primary, and in so doing we rule out a still broader class of HOD
models than the one introduced in \S\ref{subsubsec:tradhodquenching}.
A more thorough exploration of the $\mstar$-dependence of 2-halo
conformity will be presented in Paper III.

\subsection{Standard HOD Predictions of 2-Halo Conformity}
\label{subsec:twohaloconfpredhod}

As a first test, we consider the standard HOD mock described in
\S\ref{subsubsec:tradhodquenching}. For each isolated primary,
identified using the method described in \S~\ref{sec:twohaloconfmockobs}, we compute the quenched
fraction of all neighboring galaxies in spherical shells centered on
the primary. The upper panel of Fig.~\ref{fig:hodcentralconformity}
shows the mean quenched fraction as a function of the 
3D distance from the primary. Red and blue symbols correspond to the
mean quenched fractions around quenched and star-forming isolated
primaries, respectively, while the error bars reflect the uncertainties
assuming Poisson statistics.

The fact that the red and blue points are consistent with each other
within the errors indicates that this mock does not reveal any
significant 2-halo conformity over the range $1 \mathrm{Mpc/h}
\lesssim R \lesssim 5 \mathrm{Mpc/h}$. This may not come as a
surprise given that this mock, by construction, does not have any
2-halo conformity built in. However, this serves to demonstrate that satellite
contamination, as described in \S\ref{subsubsec:hodsig}, does not
introduce an artificial signal of 2-halo conformity.  
The fraction of isolated primaries that are
contaminating satellites, as opposed to true centrals, is only a few
percent.  
As is clear from the upper panel in
Fig.~\ref{fig:hodcentralconformity}, this minor level of contamination
has no detectable influence on the measured 2-halo conformity signal.

By constructing similar HOD mocks, but using different combinations of
quenching parameters $\mathrm{A}_{x}$ and $\mathrm{B}_{x}$ (see
Eqs.~\ref{eq:satellitequenching}-\ref{eq:centralquenching}), we find
that the lack of 2-halo conformity on scales $R\gtrsim1$ $\mpc$/h is robust to the particular
form of the quenching functions $\fracsqq$ and $\fraccqq$.  Hence, we conclude that satellite
contamination in the K13 measurements is unlikely to have introduced a
false signal of 2-halo conformity, and that the results of K13 are
inconsistent with models in which the quenching statistics of
galaxies are solely regulated by halo mass $\mvir.$


\begin{figure*}
\begin{center}
\includegraphics[width=.53\textwidth]{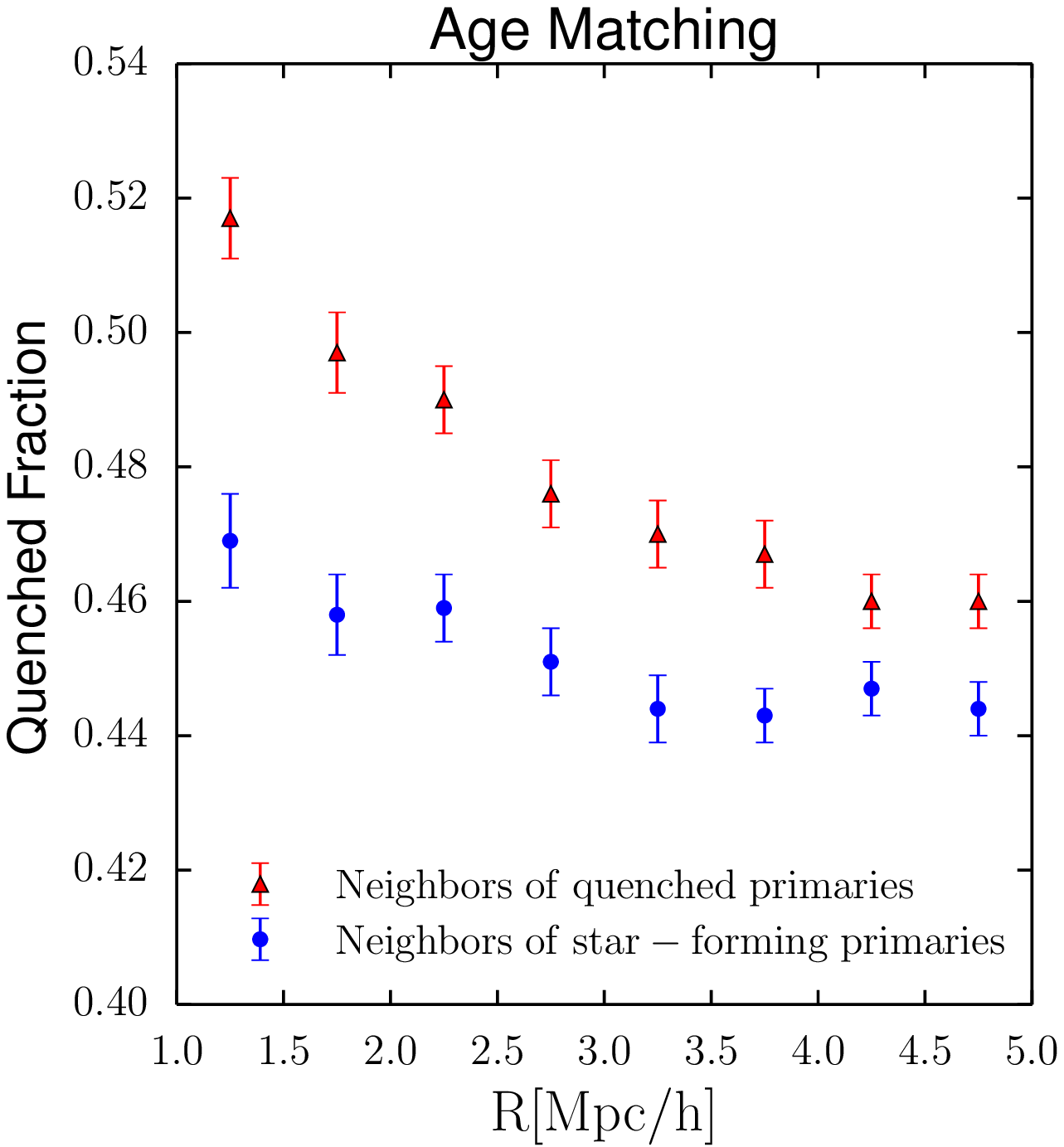}
\includegraphics[width=8cm]{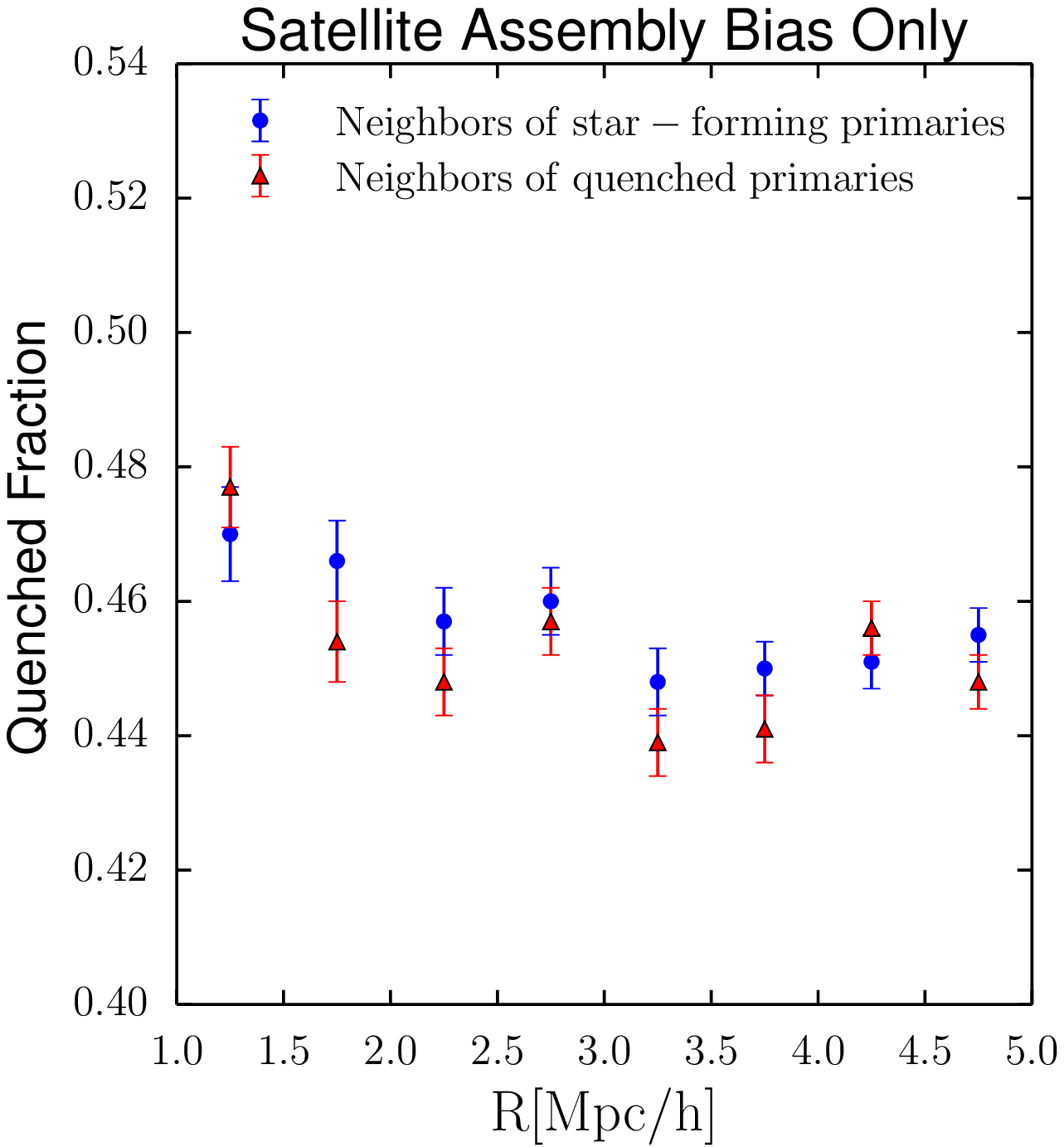}
\includegraphics[width=8cm]{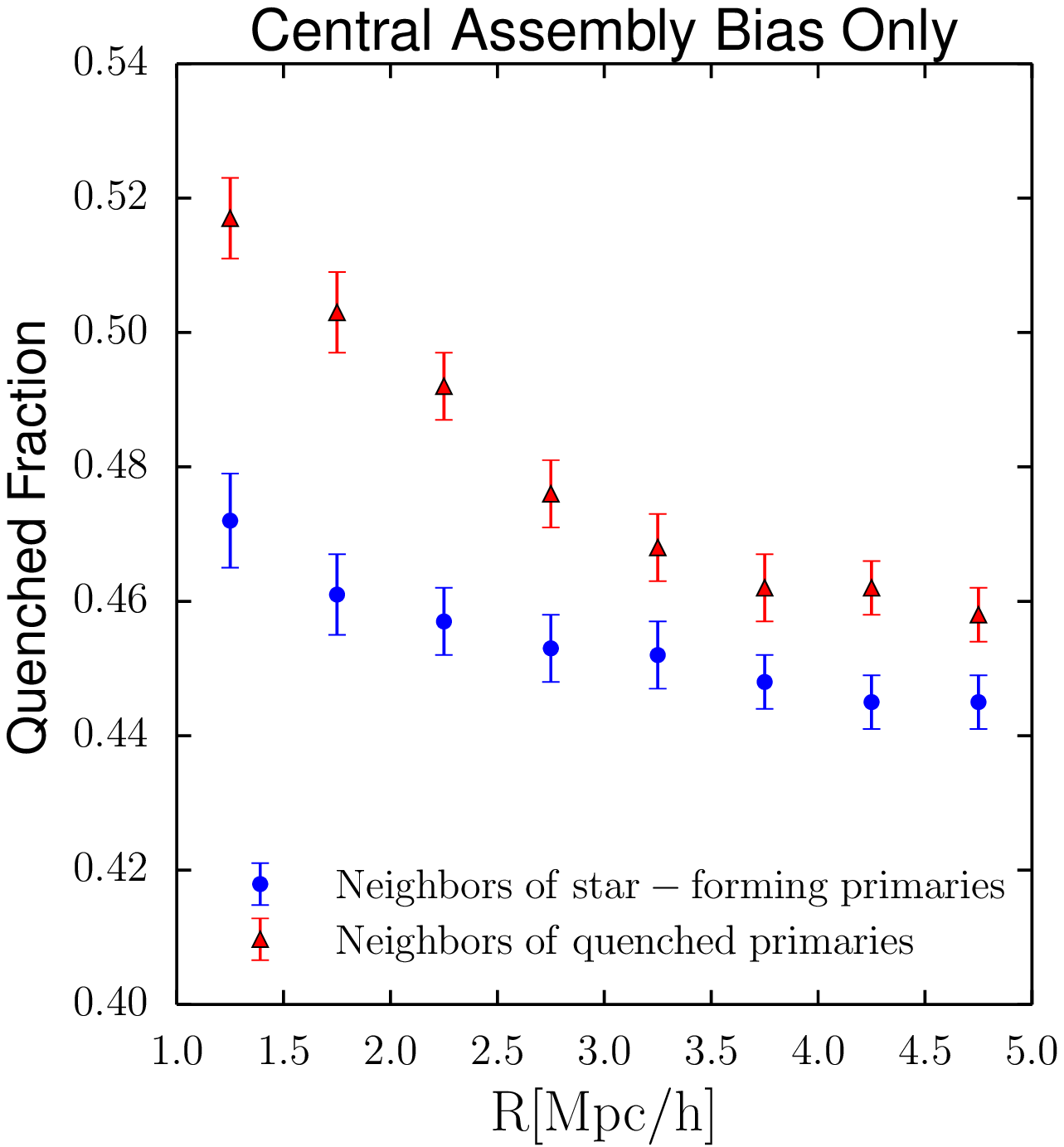}
\caption{ {\bf Age Matching prediction of 2-halo conformity.} All
  panels are the same as in Fig.~\ref{fig:hodcentralconformity}, only
  here the age matching model is used to investigate the relationship
  between galaxy assembly bias and 2-halo conformity.
  The {\em top} panel shows the 2-halo conformity prediction of age
  matching, in which galaxy assembly bias in both centrals and
  satellites is quite strong.  For the model shown in the {\em bottom
    left} panel, we have scrambled the SFRs of central galaxies
  residing in halos of similar mass $\mvir,$ which erases the strong
  signal shown in the top panel. In the {\em bottom right} panel, we
  have instead scrambled the SFRs of satellite galaxies, which only
  marginally influences the 2-halo conformity signal. Together with
  Fig.~\ref{fig:hodcentralconformity}, these results provide strong
  support for the conclusion that the level of 2-halo conformity
  measured in K13 is driven by {\em central} galaxy assembly bias (see
  \S\ref{subsec:assembias} for further discussion).  }
\label{fig:agematchingconformity}
\end{center}
\end{figure*}


\subsection{2-Halo Conformity and Satellite Backsplashing}
\label{subsec:wetzeltwohalo}

The conclusions drawn at the end of the last section raise the
following natural question: is it possible for 2-halo conformity to be
predicted by a model in which deviations from $\mvir$-determined
occupation statistics are limited to satellite galaxies? Or must
beyond-$\mvir$ effects be present in central galaxies in order to give
rise to the signal measured in K13?

The delayed-then-rapid mock is an ideal testing ground for this
purpose, because, as discussed in \S\ref{subsubsec:wetzelmodel},
galaxy assembly bias in this model is limited to satellites (including
the population of backsplash galaxies). Moreover, in this model even
the quenching of satellites is strongly influenced by (sub)halo mass: 
since satellites are presumed to evolve as centrals prior to $\tacc,$
 a satellite's ``initial'' SFR is essentially set by the mass
of its host halo at the time of infall, $\macc$ . In this sense, this
model provides a kind of minimal extension to standard halo
occupation modeling of galaxy quenching.

It is at least plausible that the treatment of satellite backsplashing
in the delayed-then-rapid model leads to some non-zero level of
2-halo conformity. As shown in \citet{wetzel_etal13},
$\sim10\%$ of the population of present-day $\mstar\sim10^{10}\msun$
centrals are actually ejected satellites residing outside the virial
radius of massive halos.  These galaxies will tend to gather near each
other through their common association with a massive host, and they
will also tend to be quenched. Thus if a large enough fraction of
these backsplashed satellites pass the K13 isolation criteria, it is
possible that the SFR correlations amongst backsplashed satellites could
produce a 2-halo conformity signal.

We directly test this possibility in the bottom panel of
Fig.~\ref{fig:hodcentralconformity}, which is the same as the top
panel but for the delayed-then-rapid mock.  The bottom panel shows
that satellite backsplashing has a negligible effect on 2-halo
conformity on scales $1 \mathrm{Mpc/h} \lesssim R \lesssim 5
\mathrm{Mpc/h}$. Recall that the conformity signal shown in
Fig.~\ref{fig:hodcentralconformity}, as with all figures in this
paper, is computed in 3D to identify the true strength of the signal.

The bottom panel of Fig.~\ref{fig:hodcentralconformity} rules out the
possibility that satellite backsplashing alone can explain the
strength of the signal reported in K13. Given the other successful,
quantitative predictions of this model, this is an interesting
observation in and of itself since it suggests that 2-halo conformity
contains information about galaxy evolution that is independent from more
traditional statistics (see \S\ref{subsec:newprobe} for further
discussion of this point). More importantly, the failure of this
mechanism to predict 2-halo conformity is highly suggestive of the
conclusion that galaxy assembly bias in centrals, not just satellites,
is responsible for the observed level of 2-halo conformity. We
investigate this possibility with the age matching mock in the
following section.

\subsection{2-Halo Conformity and Central Galaxy Assembly Bias}
\label{subsec:confassembias}

We now turn to the age matching mock, in which the prescription for
assigning SFRs to mock galaxies is predicated on the assumption that
galaxies co-evolve with their dark matter halos (see
\S\ref{subsubsec:agematching}). The top panel of
Fig.~\ref{fig:agematchingconformity} shows that, contrary to the
previous two mocks discussed above, age matching results in a highly
significant 2-halo conformity signal. 
The quenched fraction of neighboring galaxies is much larger around
quenched primaries than around star-forming primaries. We
emphasize that this result is obtained without any modification to the
age matching technique introduced in the recent trilogy of papers
describing this model \citep{HW13A,hearin_etal13b,watson_etal14}.

It is easy to understand why age matching gives rise to 2-halo
conformity: halos that collapse from the same region of the cosmic
density field have correlated assembly histories. In age matching, the
stellar mass assembly history of a galaxy is directly tied to the mass
assembly history of the galaxy's parent (sub)halo. Therefore a natural
feature of age matching is that the SFRs of galaxies in the same
large-scale environment are correlated, regardless of whether the
galaxies occupy distinct dark matter halos. 

In light of the discussion and results in \S\ref{subsec:wetzeltwohalo}, 
a natural question to ask is whether the conformity signal in age matching is driven 
by central galaxies or satellites.  In fact there are many ``beyond
$\mvir$'' phenomena predicted by age matching, including both
intra-halo effects and also large-scale effects exhibited by both
central and satellite galaxies.\footnote{See \citet{zentner_etal13} for an extensive study of
  the character of assembly bias predicted by age
  matching.}
Thus it is not clear from the top panel of
Fig.~\ref{fig:agematchingconformity} alone whether the 2-halo
conformity signal predicted by age matching is due to galaxy assembly
bias in the centrals, satellites, or both. 

Fortunately, the
following shuffling exercises make it straightforward to
identify the source of the signal in age matching. For the model
plotted in the bottom left panel of
Fig.~\ref{fig:agematchingconformity}, we start with the age matching
mock and scramble the SFR designation of {\it central} galaxies that
reside in halos of similar mass. Operationally, we bin the centrals by
the value of $\mvir$ of their host halo, using bin widths of $0.1$
dex, and randomly shuffle the SFR designations amongst the centrals in
that bin. We then recompute the 2-halo conformity signal around
isolated primaries with stellar mass $10^{10}\msun < \mstar <
10^{10.5}\msun,$ and show the result in the bottom-left panel of
Fig.~\ref{fig:agematchingconformity}. Interestingly, this scrambling
entirely erases the 2-halo conformity signal in the mock, suggesting
that central galaxies are responsible for the signal. The bottom-right
panel is the same as the bottom left, but this time we have performed
the shuffling exercise with the SFR designations of the {\it
  satellites}, leaving the central SFRs fixed to their age
matching-predicted values. The effect of this satellite scrambling is
very different: in this case, the 2-halo conformity signal is left
unadulterated. These shuffling exercises imply that the strong 2-halo
conformity signal predicted by age matching is almost entirely due to
the model's treatment of central galaxies.

To be clear, the purpose of
Figure \ref{fig:agematchingconformity} is not to claim that age matching
is the ``correct'' or only model of 2-halo conformity.  Rather, {\em the
role of the age matching model in this paper is merely to provide an
illustrative example of a galaxy-halo model in which the signal is
strong}. This is valuable, because the top panel of
Fig.~\ref{fig:hodcentralconformity} makes it highly implausible that a
standard HOD model can predict the signal, while the bottom panel
of Fig.~\ref{fig:hodcentralconformity} shows that one of the leading
empirical models of satellite quenching predicts virtually no
signal. In addition, K13 has shown that the state-of-the-art SAM of
\citet{guo_etal11b} predicts little to no 2-halo conformity signal either.
Fig.~\ref{fig:agematchingconformity} serves to demonstrate that levels
of 2-halo conformity as strong as the K13 measurements do not require
any ``spooky action-at-a-distance'', nor detailed modeling of complex
astrophysical phenomena. If the star formation history of central
galaxies traces the mass assembly history of host halos in some manner
similar to age matching, 2-halo conformity emerges naturally.

\section{DISCUSSION \& OUTLOOK}
\label{sec:discussion}

\subsection{2-Halo Conformity as an Assembly Bias Marker}
\label{subsec:assembias}

The results in \S\ref{sec:twohaloconfmockobs} support the conclusion
that the K13 measurements of 2-halo conformity constitute a detection
of {\em central galaxy assembly bias.} We use this term to mean that
central galaxy SFR is significantly correlated with some halo property
in addition to halo mass $\mvir.$ The evidence for this conclusion is
straightforward.

First, in the top panel of Fig.~\ref{fig:hodcentralconformity}, we
have shown that standard HOD models of galaxy quenching, in which
there is no assembly bias of any kind, predict that 2-halo conformity
should be zero in the stellar mass range where it has been observed
with high statistical significance ($10^{10}\msun < \mstar <
10^{10.5}\msun$). Although Fig.~\ref{fig:hodcentralconformity} only
shows that this is the case for a particular choice of HOD parameters,
we find that this conclusion holds regardless of the fiducial values
of the parameters. The K13 signal is not easily computable within the
analytical framework of the halo model, necessitating this mock
catalog-based approach.

Second, we have shown that 2-halo conformity is also zero in two very
different quenching models in which the galaxy assembly bias is
limited to satellites. The first is the delayed-then-rapid model put
forth in \citet{Wetzel_Tinker_Conroy12}, and shown in the bottom panel
of Fig.~\ref{fig:hodcentralconformity}. In the second case, we
consider an alteration to the age matching model
\citep{HW13A} wherein we scramble the SFRs of
{\em central} galaxies, at fixed halo mass. Both of
these models produce essentially zero 2-halo conformity for scales $1
\mathrm{Mpc/h} \lesssim R \lesssim 5 \mathrm{Mpc/h}$.  However, when
we then test a mock in which {\em satellite} galaxy SFRs have been
scrambled in the age matching model (bottom right panel of
Fig.~\ref{fig:agematchingconformity}), the signal is nearly as strong 
as it is in the unscrambled age matching mock (top panel of
Fig.~\ref{fig:agematchingconformity}).
This directly implies that central galaxy assembly bias is
responsible for the 2-halo conformity signal predicted by this model.

Although we acknowledge that these results do not conclusively rule
out a relationship between 2-halo conformity and satellite SFR, simply
because our exploration of satellite-based models has been far from
exhaustive, the fact that centrals dominate the abundance of galaxies
at all stellar masses of interest clearly supports the notion that 
any 2-halo conformity is likely to be driven by central galaxies rather
than satellites.

Detections of central galaxy assembly bias have been reported in a
growing body of literature. For example, numerous studies of SDSS
galaxy samples \citep[e.g.,][]{yang_etal05, yang_etal06a, wang_etal08,
  wang_etal13} have employed galaxy group finders to demonstrate that
the clustering of central galaxies (or the clustering of the groups
themselves) depends on star formation indicators even after
controlling for halo mass. The results presented in
\S\ref{sec:twohaloconf} complement these previously reported
detections with a method that has no reliance on a group-finding
algorithm: both the K13 measurements and the above group-finder based
methods support the notion that the statistics describing the
quenching of galaxies are not solely a function of halo mass $\mvir$ alone.

\subsection{2-Halo Conformity as a New Probe of Galaxy Evolution}
\label{subsec:newprobe}

The delayed-then-rapid model presented in \citet{Wetzel_Tinker_Conroy12}
attaches unique physical significance to the virial radius $\rvir$ of
the dark matter halo: central galaxy quenching statistics are
exclusively governed by the mass $\mvir$ enclosed by the virial
radius, and the SFR of satellites only begins to differ from centrals
some time after $\tacc,$ when the satellite first passes within
$\rvir$ of a larger halo.  This model can be tuned to accurately
reproduce a number of well-studied statistics based on galaxy
group catalogs, such as the quenched fraction of satellites as a
function of group mass, and the radial quenching gradients of
satellites \citep{wetzel_etal13}.  Moreover, we have shown in
Fig.~\ref{fig:modelcomparisons} that the two-point correlation
functions predicted by this model are in close agreement with the
predictions of the age matching model, which itself accurately fits
data from the SDSS \citep{HW13A, hearin_etal13b,
  watson_etal14}. Hence, we conclude that the shortcomings of the
delayed-then-rapid model are not revealed by conventional statistics,
but only become apparent when using the 2-halo conformity signal. This
demonstrates that this new statistic possesses heretofore untapped
constraining power for models of galaxy evolution.\footnote{See also
  \citet{cohn_white14} for an extensive investigation of the
  information content contained in alternative statistics to two-point
  clustering at $z \sim 0.5$.}

Another example supporting the notion that the information content of
2-halo conformity is largely independent from that of more traditional
statistics is the SAM presented in \citet{guo_etal11b}. This model
reproduces reasonably well the observed clustering of galaxies
  as a function of stellar mass and color, as well as the
distribution of galaxies within rich clusters. And yet, as shown
explicitly in K13, this model predicts little-to-no 2-halo conformity.
In K13, the authors speculate that 2-halo conformity may arise under a
modification to the Guo et al. SAM in which there is ``pre-heating''
of the inter-galactic medium (IGM) at early times. This possibility is
intriguing in light of the recent results presented in
\citet{lu_etal14}, who showed that pre-heating of the IGM may play an
important role in establishing numerous scaling relations of disk
galaxies.

\subsection{Toward a New Picture of Satellite Quenching}
\label{subsec:thenewjam}

The significant levels of 2-halo conformity measured in K13 suggest a
different picture of galaxy evolution than the one offered by 
the prevailing paradigm.  In particular, this signal
indicates that contemporary models of satellite quenching have
systematically over-estimated the significance of post-infall physical
processes on attenuating the SFRs of satellite galaxies. This
conclusion derives from the following chain of reasoning. 

First, the results of this work together with the K13 measurements
strongly suggest that the large-scale environment ($R\sim 1-5$ $\mpc$)
of a central galaxy is correlated, {\em at fixed $\mvir$}, with
processes influencing the star formation history of the central. And
so if satellites evolve as centrals for most of cosmic
time, just as subhalos lead most of
their lives as host halos, then these same processes should also be at
work in quenching the satellites.\footnote{See \citet{watson_conroy13} for further empirical
  justification of the notion that satellites do indeed evolve as
  centrals for most of cosmic history.}

Second, at fixed $\mstar$, galaxies that end up as satellites are far
more likely to evolve in a dense large-scale environment. Therefore,
the quenching mechanisms that correlate with large scale environment
at fixed $\mvir$ will have had a statistically greater influence on
the present-day satellite population, which will tend to produce
preferentially quenched satellites {\em even in the complete absence
  of post-infall specific processes}.  Models of galaxy quenching that
ignore the above fact about structure growth in CDM are left with
little choice but to rely too heavily on post-infall processes in
order to reproduce the observed excess quenched fractions of
satellites relative to centrals of the same $\mstar$.  We conclude
that careful comparisons to 2-halo conformity measurements (and other
unambiguous markers of galaxy assembly bias) should henceforth be considered an
essential component of modeling and constraining the star formation
histories of galaxies.

\section*{ACKNOWLEDGEMENTS}
\label{sec:ack}

APH is supported by the U.S. Department of Energy under contract
No. DE-AC02-07CH11359, and by a fellowship provided by the Yale Center
for Astronomy \& Astrophysics.  DFW is supported by the National
Science Foundation under Award No. AST-1202698.  Our thanks to Alyson
Brooks, Matt Becker, Risa Wechsler, Michael Busha, Yu Lu, Duncan
Campbell, and Andrew Zentner for useful discussions. Special thanks
are due to Andrew Wetzel for sharing his mock catalog, and to Andrew
Wetzel, Charlie Conroy, and Jeremy Tinker for extensive comments on an
early draft. We thank The Sound Defects for {\em Iron Horse.}

\bibliography{./bhm1.bib}

\appendix

\section{The (In)Significance of Distinct $\mstar-\mvir$ Relations for
Star-Forming and Quenched Centrals}
\label{subsubsec:smhm}

As discussed in \S\ref{subsec:twohaloconfpredhod}, our standard HOD
mock does not reveal any significant 2-halo conformity, indicating
that satellite contamination is not a major concern. It also suggests
that 2-halo conformity is inconsistent with the standard
``$\mvir$-only'' paradigm for halo occupation statistics. However, the
HOD model that we considered implicitly assumes that the
$\mstar-M_{\mathrm{h}}$ relation is the same for quenched and
star-forming centrals. Yet, results deriving from the analysis of
satellite kinematics indicate that this assumption may be violated,
and that the average halo mass of star-forming central galaxies may be
distinct from that of quenched centrals \citep{more10}. This is also
supported by HOD analyses of two-point clustering and galaxy-galaxy
lensing \citep{mandelbaum06b, tinker_etal13}.  Naively, this indicates
that the range of models spanned by the quenching formulation
presented in \S\ref{subsubsec:tradhodquenching} may not be general
enough to rule out ``$\mvir$-only'' models of galaxy quenching
statistics.

In fact, different $\mstar-M_{\mathrm{h}}$ relations for star-forming
and quenched centrals could potentially produce a 2-halo conformity
signal. Suppose that quenched centrals reside, on average, in more
massive halos than star-forming centrals of the same stellar
mass. Since halo clustering strength increases with halo mass,
quenched centrals will find themselves surrounded by more massive
halos than star-forming centrals. And because the quenched fraction
increases with halo mass for centrals and satellites alike, the
environment of quenched centrals will tend to be more quenched than
that of star-forming centrals. Hence, it seems plausible that halo
occupation statistics in which quenched and star forming centrals have
different $\mstar-M_{\mathrm{h}}$ relations could give rise to 2-halo
conformity. However, as we demonstrate below, the strength of the
effect turns out not to be significant.

\begin{figure}
\begin{center}
\includegraphics[width=8cm]{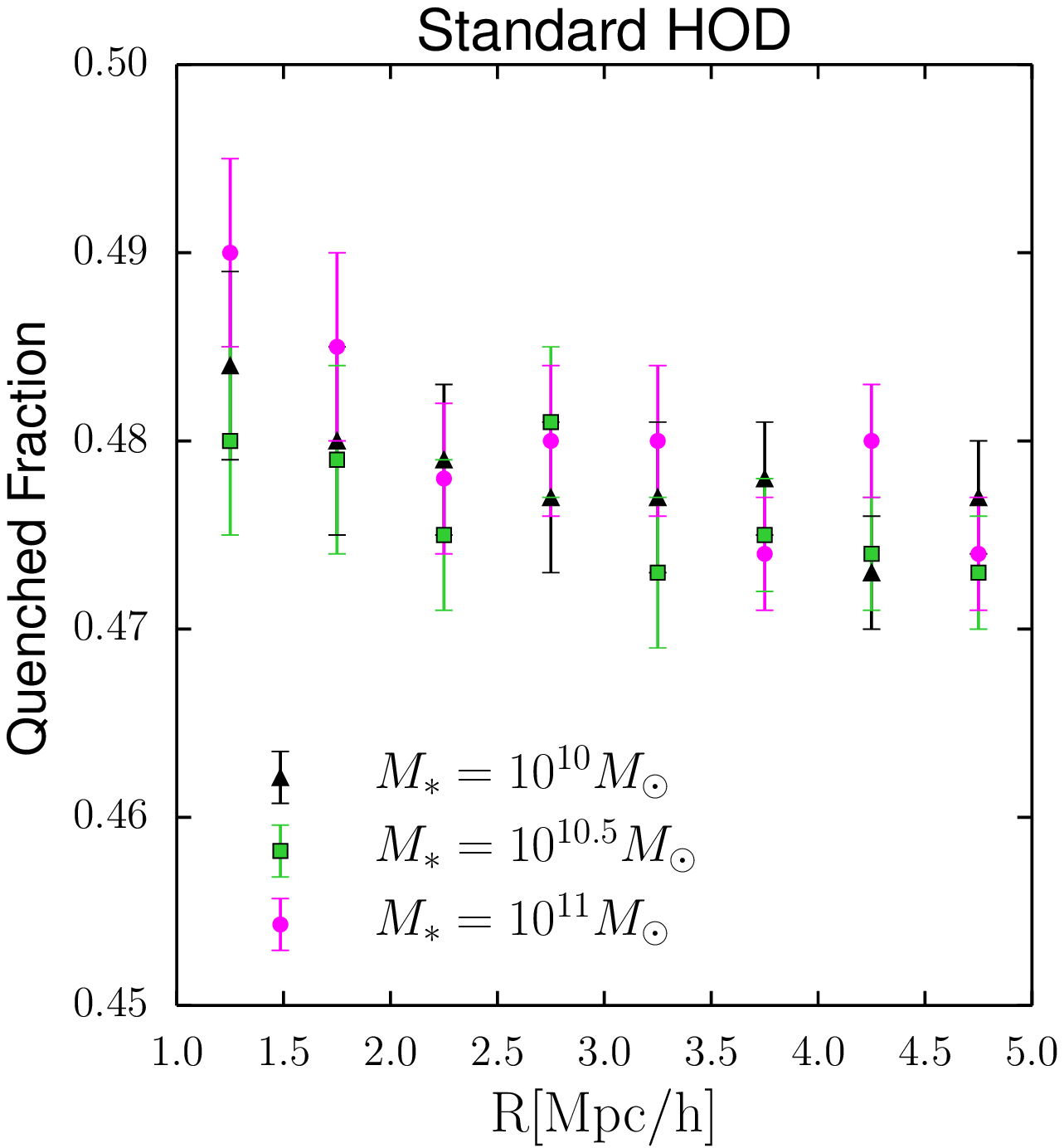}
\caption{The quenched fraction around isolated primary
  samples of different stellar mass. Isolated primaries have {\em not} been split 
   into star-forming and quenched subpopulations. This figure
  shows that the shortcoming of the HOD shown in the top panel of
  Fig.~\ref{fig:hodcentralconformity} cannot be resolved by allowing
  quenched and star-forming populations to have distinct
  $\mstar-M_{\mathrm{halo}}$ relations. }
\label{fig:smdependence}
\end{center}
\end{figure}

Rather than constructing a mock in which we use different
$\mstar-M_{\mathrm{h}}$ relations for quenched and star forming
centrals, we demonstrate the inability of different
$\mstar-M_{\mathrm{h}}$ relations to induce 2-halo conformity by
comparing the quenched fractions around isolated primaries of
different stellar mass. The results, obtained using our standard HOD
mock (see \S~\ref{subsubsec:tradhodquenching}), are shown in Fig.~\ref{fig:smdependence}, where different
symbols correspond to different stellar mass bins, as indicated. The
quenched fractions shown are for the entire samples (star-forming plus
quenched) of isolated primaries. As is evident, the quenched fraction
around isolated primaries correlates very weakly with the stellar mass
of the primary. There is a hint for some dependence on small scales
($R < 1.5\mathrm{Mpc}/h$), but this is barely significant. Hence, even if
quenched and star-forming centrals were to have different
$\mstar-M_{\mathrm{h}}$ relations, this would not introduce
significant 2-halo conformity. This is easy to understand: in the
range of stellar mass considered here, the corresponding host halos
are less massive than the redshift-zero collapse mass, which implies
that halo bias is only a weak function of halo mass. Based on these
findings, and on those described in the main text, we conclude that
{\it the 2-halo conformity signal detected by K13 in SDSS data highlights
  a generic failure of halo occupation models in which the quenching
  statistics are governed by halo mass alone}.


\end{document}